\documentstyle[12pt]{article}
\def \be{\begin{equation}}
\def \ee{\end{equation}}
\def \ba{\begin{array}{l}}
\def \ea{\end{array}}
\def \a{\alpha}

\def \G{\Gamma}
\def \d{\delta}
\def \D{\Delta}
\def \e{\epsilon}
\def \f{\phi}

\def \t{\tau}

\def \lm{\lambda}
\def \n{\nabla}
\def \p{\varphi}

\def \ol{\overline}

\def \la{\langle}
\def \ra{\rangle}
\def \[{\left[}
\def \]{\right]}
\def \({\left(}
\def \){\right)}
\def \Tr{{\rm Tr}}
\def \R{R_{c}}
\def \T{T_{c}}
\def \fr{\frac}
\def \I{\int d^{D}x}
\def \2{\frac{1}{2}}
\def \4{\frac{1}{4}}
\def \lb{\label}
\newcommand{\sectio}[1]{\section{#1}\setcounter{equation}{0}}

\begin{document}
\frenchspacing
\setlength{\parskip}{2mm}


\begin{center}

{\Large \bf Griffith Singularities and the Replica Instantons in the Random Ferromagnet}
\vskip .4in
Vik. S. Dotsenko
\vskip .2in
Laboratoire de Physique Theorique de  l'Ecole Normale Superieure,\\
24, rue Lhomond, 75231 Paris Cedex 05, France\\
and\\
L. D. Landau Institute for Theoretical Physics,\\
Russian Academy of Sciences,\\
Kosygina 2, 117940 Moscow, Russia\\

\end{center}
\vskip 1in

\begin{abstract}
The problem of existence of non-analytic (Griffith-like) contributions to the free
energy of weakly disordered Ising ferromagnet is studied from the point of view
of the replica theory. The consideration is done in terms of the usual
random temperature Ginzburg-Landau Hamiltonian in space dimensions $D < 4$
in the zero external magnetic field.
It is shown that  in the paramagnetic phase, at temperatures not too
close to $T_{c}$ (where the behaviour of the pure system is correctly described by
the Gaussian approximation),  the free energy of the system has additional 
non-perturbative contribution of the form $\exp\{ -(const) \frac{\tau^{(4-D)/2}}{u} \}$
(where $\tau = (T-T_{c})/T_{c}$),  which has essential singularity in the parameter $u \to 0$ 
which describes the strength of the disorder. It is demonstrated that this contribution
appears due to non-linear localized (instanton-like) solutions of the 
mean-field stationary equations which are characterized by the special type of the 
replica symmetry breaking . 
It is argued that physically these replica instantons describe the contribution
from rare spatial "ferromagnetic islands" in which local (random) 
temperature is below $T_{c}$.
\end{abstract}

\newpage

\sectio{Introduction}

According to the original statement of Griffith \cite{grif}, the free energy of the
random Ising ferromagnet in the temperature interval above its
ferromagnetic phase transition point $\T$ and below the critical point 
$\T^{(0)}$ of the corresponding pure system must be a non-analytic function 
of the external magnetic field $h$, such that in the limit $h \to 0$ the free energy 
as a function of $h$ have essential singularity. 
Since this type of phenomena, namely, existence of non-analytic non-perturbative 
contributions to thermodynamical functions in random systems, seems to be 
rather general one, at present it became common to call any such contribution 
as the "Griffith singularity".

Due to intensive theoretical \cite{grif-t} and numerical \cite{grif-n} studies of the 
Griffith singularities it was also discovered that the {\it dynamical} properties of 
the system in the temperature interval $\T<T<\T^{(0)}$ 
are not just ordinary paramagnetic. In particular, the time correlation functions
here are described by the so-called stretched-exponential asymptotic behaviour
which is much slower than the usual exponential one, as it should be
in the paramagnetic phase.  To underline that the properties of the system in the 
temperature interval $\T < T < \T^{(0)}$ are not quite paramagnetic, it became common
to call the state of the system here as the "Griffith phase". 

At the level of "hand-waving arguments" the dynamical Griffith phenomena
can be explained "theoretically" rather easily: considering e.g.
the bond diluted Ising model, one can note that at temperatures below $\T^{(0)}$ 
in the "ocean" of the zero magnetization paramagnetic background the random system 
must contain disconnected locally ordered "ferromagnetic islands" (composed 
only of the pure system bonds) of all sizes, which, in turn, creates the whole
spectrum (up to infinity) of relaxation times. Having an infinite spectrum of
relaxation times, with some imagination, it is not difficult to derive any
relaxation law one likes, and the stretched-exponential one in particular. 

Although it is commonly believed that the main point of the above "explanation", 
namely the existence of infinite number of local minima states,  must be a general key 
for understanding the Griffith phenomena (both dynamical and statistical mechanical), 
despite many efforts during last thirty years, it turned out to be extremely difficult 
to construct more or less elaborated and convincing theory. For that reason any 
progress in understanding of the effects produced by numerous local minima states 
(which, so to say, are away from the perturbative region) looks valuable.

In this paper I am going to study non-perturbative contributions to the thermodynamical
functions  of weakly disordered (random temperature)
$D$-dimensional ($D < 4$) Ising ferromagnet in the paramagnetic phase away 
from the critical point. In the continuous limit this system can be described by the 
usual Ginsburg-Landau Hamiltonian:

\be
\lb{g1}
H\[\f(x);\d\t(x)\] = \I \[ \2 \(\n\f(x)\)^{2} + \2 \(\t - \d\t(x)\) \f^{2}(x) +
 \4 g \f^{4}(x) \] 
\ee
Here $\t \equiv (T - \T)/\T \ll 1$ is the reduced temperature, and the quenched disorder 
is described by random spatial fluctuations of the local transition temperature $\d\t(x)$ 
whose probability distribution is taken to be symmetric and Gaussian:

\be
\lb{g2}
P[\d\t] = p_{0} \exp \Biggl( -\frac{1}{4u}\I (\d\t(x))^{2} \Biggr) \ ,
\ee
where $u \ll g$ is the small parameter which describes the strength of the disorder,
and $p_{0}$ is irrelevant normalization constant. 
For notational simplicity, I define the sign of $\d\t(x)$ in eq.(\ref {g1}) so that
positive fluctuations lead to locally ordered regions, whose effects will be
the object of our further study.

As far as the corresponding pure system ($u \equiv 0$) is concerned, it is well known that
in the close vicinity of $\T$,  at $|\t| \ll \t_{g} \sim g^{2/(4-D)}$, its properties
are defined by non-Gaussian critical fluctuations (which can be
studied e.g. in terms of the $\e$-expansion renormalization group approach), while
away from $\T$, at $|\t| \gg \t_{g}$, the situation is getting Gaussian, and
everything becomes very simple. Here the total magnetization of the system is defined
by the order parameter $\la \f\ra \equiv \f_{0}(\t)$ which is equal to 0 above $ \T$,
and it is equal to $\pm \sqrt{|\t|/g}$ below $\T$; the asymptotic behaviour of the 
correlation function $G(x-x') \equiv (\la\f(x)\f(x')\ra - \f_{0}^{2}) $ is defined only
by the Gaussian fluctuations: $G(x) \sim |x|^{-(D-2)}$; and the singular part of the
free energy $f(\t)$ scales with the temperature as $f(\t) \simeq \t^{D/2}$.

Usually, the random system, defined by the Hamiltonian (\ref{g1}), were studied  from the
point of view of the effects produced by the quenched disorder on the critical
phenomena in the close vicinity of the phase transition point. Renormalization group
consideration shows that if the temperature is not too close to $\T$, at 
$\t_{u} \ll \t \ll \t_{g}$ (where the disorder dependent crossover temperature
scale $\t_{u} \sim u^{1/\a}$ is defined by the specific heat critical exponent $\a > 0$ of the 
pure system) the critical behaviour is essentially controlled by the pure system
fixed point, and the disorder produces only irrelevant corrections. On the other hand,
in the close vicinity of the critical point, at $\t \ll \t_{u}$, the critical behaviour moves
into a new universality class defined by the so called random fixed point, which turns
out to be universal \cite{crit1}. In recent years, however, this very nice physical 
picture has been questioned on the grounds that the renormalization group
approach completely misses the presense of numerous local minima configurations
of the random Hamiltonian (\ref{g1}), which, in principle, may cause the
spontaneous replica symmetry breaking in the interaction parameters of the 
critical fluctuations, which, in turn, may ruin the above physical scenario \cite{d-rsb}.

Leaving the discussion of this very difficult problem for future analycis, 
in this paper I would like to pose much more simple question: how  
the thermodynamic functions of this system depend on the strength of the disorder $u$ 
(in the limit $u \to 0$) far way from $\T$, at $\t \gg \t_{g}$, where the behaviour of the 
pure system is correctly described by the Gaussian approximation? 
It tuns out that even this, as if almost trivial question is not so easy to answer.

Of course, first of all, one can proceed in a straightforward way, developing 
the perturbation theory in powers of the parameter $u$ at the background
of the pure system paramagnetic state $\la\f\ra = 0$ using the Gaussian
approximation for the thermal fluctuations. There is nothing wrong in this
approach, but the problem is that it can not give {\it all} thermodynamic 
contributions which exist at $u \not= 0$. The drawback of this type of the
perturbation theory is the same as that of the renormalization group: it completely
misses the existence  of numerous (macroscopic number) local minima
configurations of the random Hamiltonian (\ref{g1}). 

At the level of "hand waving
arguments" it is very easy to see what all these off-perturbative states are.
At any $u \not= 0$ there exists finite (exponentially small) density of "ferromagnetic 
islands" in which local (random) temperature is below $\T$ (such that
$\d\t(x) > \t$), and the minimum energy configurations here are achieved at
non-zero local value of the order parameter: $\f_{0}(x) \sim \pm\sqrt{(\d\t - \t)/g}$.
Since the spatial density of such islands is finite, and each island provide
two ($\pm$) possibilities for the local magnetization,  the total number of
the local minima configurations in the system must be exponential in its volume.

Formally, to take into account the contributions of all these states, one has to
proceed as follows. For an arbitrary quenched function $\d\t(x)$ one has to find
all possible local minima solutions of the saddle point equation:

\be
\lb{g3}
-\D\f(x) +(\t - \d\t(x))\f(x) + \f(x)^{3} = 0
\ee
Then one has to substitute these solutions into the Hamiltonian (\ref{g1}) and 
calculate the corresponding thermodynamic weights. Next, to
compute the partition function one has to perform summation over all the solutions, 
and finally to get the corresponding free energy one has to take the logarithm of the 
partition function and average it over random functions $\d\t(x)$ with the probability
distribution (\ref{g2}). Clearly, it is hardly possible that such a programme
can be implemented. 

On the other hand, as usual, for the systems which contain quenched disorder 
we can use the standard replica method and reduce the problem of the quenched 
averaging to the annealed one for $n$ copies of the original system:

\be 
\lb{g4}
F = - \ol{\(\ln Z\)} = - \lim_{n\to 0} \fr{1}{n} \( \ol{Z^{n}} - 1 \)
\ee
where $\ol{(...)}$ denotes the averaging over random functions $\d\t(x)$ 
with the probability distribution (\ref{g2}), and 

\be
\lb{g5}
Z[\d\t(x)] = \int {\cal D}\f(x) \exp\( -H\[\f(x);\t(x)\] \)
\ee
is the disorder dependent partition function which is given by the functional integration 
over configurations of the field $\f(x)$. 

Simple Gaussian integration over $\d\t(x)$ in eq.(\ref{g4}) yields:

\be
\lb{g6}
 \ol{Z^{n}} = \prod_{a=1}^{n}\[\int {\cal D}\f_{a}(x)\] \exp\( -H_{n}[\f_{a}(x)] \)
\ee
where

\be
\lb{g7}
\ba
H_{n}[\f_{a}(x)] = \\
\\
= \I \[ \2 \sum_{a=1}^{n}\(\n\f_{a}\)^{2} + \2 \t \sum_{a=1}^{n} \f_{a}^{2} +
 \4 g \sum_{a=1}^{n} \f_{a}^{4} -\4 u \sum_{a,b=1}^{n}  \f_{a}^{2} \f_{b}^{2}\]
\ea
\ee
is the {\it spatially homogeneous} replica Hamiltonian. 

Now, if we are intended to take into account non-trivial local minima states, instead of
solving the original inhomogeneous stationary equation (\ref{g3}), we can consider
the corresponding replica saddle-point equations:

\be
\lb{g8}
-\D\f_{a}(x) + \t \f_{a}(x) + \f_{a}^{3}(x) - u \f_{a}(x)\sum_{b=1}^{n}  \f_{b}^{2}(x) = 0
\ee
Since until now all the transformations were exact, these equations must contain
(may be in a slightly hidden way) all the relevant non-trivial states which
in the language of the original random Hamiltonian correspond to rare ferromagnetic
islands. 

At this stage we can note one very simple point. Looking for various types of solutions
of the above equations one can try first of all the simplest possible "replica symmetric"
ansatz, in which the fields in all replicas are assumed to be equal: $\f_{a}(x) = \f(x)$.
In this case the last term in the eqs.(\ref{g8}) (which contains the factor
$\sum_{b=1}^{n}  \f_{b}^{2}(x) = n \f^{2}(x)$) drops away in the limit $n \to 0$, and
these equations reduce to the {\it pure system} saddle-point equation:

\be
\lb{g9}
-\D\f(x) + \t \f(x) + \f(x)^{3} = 0
\ee
which at $\t > 0$ has only trivial solution $\f(x) \equiv 0$. It means that
in any non-trivial solution of the eqs.(\ref{g8}) the fields $\f_{a}(x)$ in different
replicas {\it can not} be all equal. In other words, the symmetry among replicas
in the {\it replica vector} $\f_{a}(x)$ must be broken. 

The methodological aspects of how to handle with the vector replica symmetry breaking
situation in various disordered systems are described in the paper \cite{dm}.
In the next Section this method will be applied for the problem described above. 
It will be shown that indeed, in the high-temperature region ($\t > 0$) eqs.(\ref{g8}) 
have non-trivial localized 
(having finite size and finite energy) solutions in which the replica symmetry
in the fields $\f_{a}(x)$ is broken. The formal summation over all such solutions 
provides the contribution to the free energy of the typical Griffith-like form:
$\exp\{ -(const) \frac{\tau^{(4-D)/2}}{u} \}$. It will also be shown that the
mean-field approach (in which the critical fluctuations are ignored) used in this paper 
is grounded only if the temperature is not to close to $\T$, namely at
$\t \gg \t_{g} \sim g^{2/(4-D)}$, the same as in the classical Ginsburg-Landau theory.
Finally, it will be demonstrated how this type of non-analytic contribution
to the free energy can be estimated from purely physical arguments
taking into account probabilities for the typical "ferromagnetic islands".

To avoid possible misunderstandings, 
in conclusion of this introductory Section I would like to note the following essential
point. The problem considered in this paper is actually rather far from the original 
one studied by Griffith as well as by many other people later on.  
Since the shift of $\T$ in the weakly disordered ferromagnet compared to $\T^{(0)}$ of 
the pure
system is of the order of $\sqrt{u}$, in the limit $u \ll g$ the interval of temperatures 
$\T < T < \T^{(0)}$ where the so called Griffith phase is expected to take place, 
appears to be well inside of the temperature interval $\t_{g} \sim g^{2/(4-D)}$
where the critical fluctuations are essential, and where the mean field approach
considered in this paper can not be used. For that reason, in the considered range of 
temperatures $\t \gg \t_{g}$ it is hardly reasonable to look for non-analytic behaviour 
of the free energy as the function of the external magnetic field (at least the present
approach in terms of the replica instantons modified by the external field $h$ does 
not seem to indicate on any non-analyticity in $h$). 
The aim of this paper is just to demonstrate that in additional to the "usual"
Griffith singularities in terms of the external field, the free energy of the random
ferromagnet (in the zero magnetic field) must also be non-analytic in the value of the 
parameter which describes the strength of the disorder.

\sectio{Replica Instantons}

Following the general strategy developed in the the paper \cite{dm},
let us assume that in addition to the trivial replica symmetric (RS) solutions
of the saddle-point equations (\ref{g8}) there exist other types of solutions,
which are {\it well separated} in the configurational space from the RS state.
In this case, denoting the contribution of these non-trivial states
by the lable "replica symmetry breaking" (RSB), the replica partition function,
eq.(\ref{g6}), can be decomposed into two parts:

\be
\lb{g10}
\ol{Z^{n}} = Z_{RS} + Z_{RSB}
\ee
where $Z_{RS}$ contains all the perturbative contributions in the vicinity of the
trivial state $\f_{a}(x) = 0$. As usual, this partition function can eventually be
represented in the form:

\be
\lb{g11}
Z_{RS} = \exp\( - n V f_{RS}\)
\ee
where $V$ is the volume of the system, and $f_{RS}$ is the free energy density,
which contains the pure system leading term $\sim \t^{D/2}$ (at temperatures
not too close to $\T$, $\t \gg \t_{g}$), plus the perturbation
series in powers of the disorder parameter $u$. 

Thus, in terms of the general replica approach, according to eq.(\ref{g4})
for the total free energy we get:

\be
\lb{g12}
F = V f_{RS} + F_{RSB}
\ee
where the additional RSB part of the free energy

\be
\lb{g13}
F_{RSB} = - \lim_{n\to 0} \fr{1}{n} Z_{RSB}
\ee
must contain all non-perturbative contributions (if any) which are away
from the trivial state $\f_{a} = 0$. It is this part of the free energy which
will be point of our further study.

The simplest possible non-trivial replica structure for the solutions
of the saddle-point equations (\ref{g8}) can be taken in the following
form (see \cite{dm}):

\be
\lb{g14}
\f_{a}(x) = \left\{ \begin{array}{ll}
                 \f(x) & \mbox{for $a = 1, ..., k$}\\
                 0     & \mbox{for $a = k+1, ..., n$}
                          \end{array}
                 \right.
\ee
where $k$ is the integer value parameter: $k = 1, 2, ..., n$ which defines a given
structure of the trial replica vector $\f_{a}$  (note that the value $k = 0$ should be
excluded since it describes the trivial RS solution which is already taken into 
account in $f_{RS}$).

Substituting this anzatz into eqs.(\ref{g8}) as well as into the replica Hamiltonian (\ref{g7}), 
one finds that for a given value of the parameter $k$ the fields $\f(x)$ in eq.(\ref{g14}) 
are defined by the solutions of the following saddle-point equation:

\be
\lb{g15}
-\D\f(x) + \t \f(x) - \lm(k) \f(x)^{3} = 0
\ee
and the thermodynamic weight of any such solution is defined by the energy: 

\be
\lb{g16}
E(k) = k \I \[ \2 \(\n\f(x)\)^{2} + \2 \t \f^{2}(x) - \4 \lm(k) \f^{4}(x) \] 
\ee
where

\be
\lb{17}
\lm(k) = (uk  - g)
\ee
Summing over the parameter $k$ and taking into account the combinatoric factor 
which is the number of permutations among replicas in the ansatz structure 
(\ref{g14}) for the free energy, eq.(\ref{g13}), one gets:

\be 
\lb{g18}
F_{RSB} = - \lim_{n\to 0} \fr{1}{n} \sum_{k=1}^{n} \fr{n!}{k! (n-k)!} \exp\{ -E(k)\}
\ee
To take the limit $n \to 0$ the series in the above 
equation can be be represented as follows:

\be
\lb{g19}
F_{RSB} = -\lim_{n\to 0}\fr{1}{n}
\sum_{k=1}^{\infty}\fr{\G(n+1)}{\G(k+1)\G(n-k+1)}
\exp\{ -E(k)\}
\ee
Here the summation over $k$ is extended beyond $k=n$ to $\infty$ since the
gamma function is equal to infinity at negative integers.
Now using the relation $\G(-z) = \pi [z\G(z)\sin(\pi z)]^{-1}$, we can perform 
the analytic continuation $n \to 0$:

\be
\lb{g20}
\ba
\fr{\G(n+1)}{\G(k+1)\G(n-k+1)}=
\\
\\
=\fr{\G(n+1) (k-1-n) \G(k-1-n) \sin(\pi(k-1-n))}{\pi \G(k+1)}|_{(n \to 0)} \; \;
\simeq n \fr{(-1)^{k-1}}{k}
\ea
\ee
Thus, for the free energy (\ref{g18}) one obtains:

\be 
\lb{g21}
F_{RSB} = -  \sum_{k=1}^{\infty} \fr{(-1)^{k-1}}{k}  \exp\{ -E(k)\}
\ee

At this stage we can note the following important point. For any {\it non localized}
(e.g. space-independent) solution,
such that its energy (\ref{g16}) is divergent with the volume $V$ of the system,
the corresponding contribution to the free energy (\ref{g21}) will not be proportional
to $V$, but instead, it will contain the volume in the exponential
factor. It means that at least for the bulk properties of the system this type of solutions  
must be irrelevant.

Thus, we have to look for  {\it localized} solutions: the ones which are 
local in space (breaking translation invariance) and which have {\it finite} 
(volume-independent) energy. Let us suppose that such instanton-type 
solutions do exists (see below), and that for a given $k$ the solution is characterized 
by the spatial size $R(k)$. Then, if we take into account only one-instanton contribution
(or in other words if we consider a gas of {\it non-interacting} instantons),
due to obvious entropy factor $V/R^{D}$ (which is the number of positions
of the object of the size $R$ in the volume $V$) we  get the free energy  proportional 
to the volume:

\be
\lb{g22}
F_{RSB}  \simeq - V \sum_{k=1}^{\infty} \fr{(-1)^{k-1}}{k} R^{-D}(k) \exp\{-E(k)\}
\ee

Now let us come back to the saddle-point equation (\ref{g15}), and let us consider the
range of the parameter $k$ such that $\lm(k) = (uk - g) > 0$ (i.e. $k > [g/u]$). 
Rescaling the fields:

\be
\lb{g23}
\f(x) = \pm \sqrt{\fr{\t}{\lm(k)}} \psi(x\sqrt{\t})
\ee
instead of eq.(\ref{g15}) one get the following differential equation which 
contains no parameters:

\be
\lb{g24}
-\D \psi(z) + \psi(z) - \psi^{3}(z) = 0
\ee
Correspondingly, for the energy, eq.(\ref{g16}), one obtains:

\be
\lb{g25}
E(k) = \fr{k}{uk - g} \t^{(4-D)/2} E_{0}
\ee
where

\be
\lb{g26}
E_{0} = \int d^{D}z \[ \2 (\n\psi(z))^{2} + \2 \psi^{2}(z) - \4 \psi^{4}(z) \]
\ee

The equation (\ref{g24}) is well know in the field theory (see e.g. \cite{inst}):
it is for  the present choice of signs of the linear and the cubic terms (which imposes 
the conditions:  $\t > 0$ and $k > [g/u]$) in dimensions $D < 4$  this equation has 
spherically symmetric instanton-like solutions such that:

\be
\lb{g24a}
\ba
\psi(|z| \leq 1) \simeq \psi(0) \sim 1, \\
\\
\psi(|z| \gg 1) \sim \exp (-|z|) \to 0. 
\ea
\ee
The energy, eq.(\ref{g26}), 
of such a solution is a finite and {\it positive} number.
Of course, for a {\it generic} value of the field $\psi(0)$ at the origin, the solution
tends to the values $\psi(|z| \to \infty) = \pm 1$ which are the extrema of the
potential $\2\psi^{2} - \4\psi^{4}$, and any such solution has divergent
with volume energy (\ref{g26}). However, there exists a discrete set of initial
values $\psi_{0}$ for which the solutions (exponentially) tends to zero at infinity,
and which have finite energies. It can be show that the solution with the minimal energy
$E_{0}$ corresponds to the minimal value of $|\psi_{0}|$ in the set. In particular,
at $D = 3$, $\psi_{0} \simeq 4.34$ and $E_{0} \simeq 18.90$.
For our further calculations with the exponential accuracy it will be sufficient to take into 
account only the solution with the minimal energy. 

According to the rescaling (\ref{g23}), in terms of the original fields $\f(x)$ the size 
of the instanton  is $R = \t^{-1/2}$ (note that it does not depends on $k$), which coincides 
with the usual correlation length of the Ginsburg-Landau theory. 
Substituting this value of $R$ as well as the energy (\ref{g25}) of the instanton into 
the series (\ref{g22}) for the free energy one gets:

\be 
\lb{g27}
F_{RSB}  \simeq 
- V \t^{D/2} \sum_{k > [g/u]}^{\infty} \fr{(-1)^{k-1}}{k} 2^{k} 
\exp\[-E_{0} \fr{k}{uk-g} \t^{(4-D)/2} \]
\ee
(the factor $2^{k}$ appears due to independent summation over $\pm$ signs,
eq.(\ref{g23}), in $k$ non-zero replicas, eq.(\ref{g14})).
It can be easily shown that under the considered conditions on the parameters $u, g$ 
and $\t$ ($u \ll g \ll 1$, and $ g^{2/(4-D)} \ll \t \ll 1$) the leading contribution in the 
above series with the exponential accuracy comes from the region $k \gg g/u \gg 1$:

\be 
\lb{g28}
\fr{1}{V} F_{RSB}  \simeq  \t^{D/2} \exp\[-E_{0} \fr{\t^{(4-D)/2}}{u} \] \times
\sum_{k \gg g/u}^{\infty} \fr{(-1)^{k-1}}{k} 2^{k}  
\ee
Here the absolute value of the series 
$\sum_{k = k_{o}\gg 1}^{\infty} k^{-1} (-1)^{k-1} 2^{k}$ can be
estimated by the upper bound $\sim k_{o}^{-1} 2^{k_{o}}$, and  since it is assumed that
$\t \gg g^{2/(4-D)}$  the term $\fr{g}{u}\ln 2$, which appear in the exponential, can be 
dropped in comparison with $E_{0}\fr{\t^{(4-D)/2}}{u}$. Thus, for the density of the free energy
we finally obtain the following contribution:

\be
\lb{g29}
\fr{1}{V} F_{RSB} \sim \exp\[-E_{0} \fr{\t^{(4-D)/2}}{u} \]
\ee
(where we drop all pre-exponential factors, which within the present accuracy of 
calculations can not be defined).
 
\sectio{Fluctuations}

Note first of all, that one should not be confused by the "wrong" sign of the
$\f^{4}$ interaction term in the energy function (\ref{g16}), which for the usual
field theory would indicate on its absolute instability.
Here, as usual in the replica theory, in the limit $n\to 0$ everything turns
"up down", so that the minima of the physical free energy actually correspond to 
the {\it maxima}
of the replica free energy. It can be easily shown (see below)
that formal integration over $n$-component replica fluctuations around
considered instanton solution in the limit $n\to 0$ yields physically 
sensible result.

Proceeding the same way as in the usual Ginsburg-Landau theory, let us determine 
under which conditions the above mean-field approach used to derive the result (\ref{g29}) 
can be valid. Introducing small fluctuations $\p_{a}(x)$ near the instanton solution,
eqs.(\ref{g14}) and (\ref{g23}): $\f_{a}(x) = \f^{(inst)}_{a}(x) + \p_{a}(x)$, 
in the Gaussian approximation we get the following Hamiltonian
for the fluctuating fields:

\be
\lb{g30}
H[\p] =  \I \[ \2 \sum_{a=1}^{n} \(\n\p_{a}(x)\)^{2} + 
\2 \t \sum_{a,b=1}^{n} T_{ab}(x) \p_{a}(x)\p_{b}(x)  \] 
\ee
where the matrix $T_{ab}(x)$ contains  the $k\times k$ block:

\be
\lb{g31}
T^{(k)}_{ab}(x) = \(1 - \fr{uk - 3g}{uk - g} \psi^{2}(x\sqrt{\t})\)\d_{ab} - 
                           \fr{2u}{uk-g} \psi^{2}(x\sqrt{\t})
\ee
(where $a,b =1, ..., k$) and the diagonal elements for the remaining $(n-k)$ replicas:

\be
\lb{g32}
T^{(n-k)}_{ab} = \(1 - \fr{uk}{uk-g} \psi^{2}(x\sqrt{\t}) \) \d_{ab}
\ee
(where $a,b = k+1, ..., n$). 
Here the function $\psi(z)$ is the instanton solution, eq.(\ref{g24a}).

Since the mass term in the Hamiltonian (\ref{g30}) is proportional to $\t$,
the behaviour of the correlation function of the fluctuating fields
at scales \\
$|x| \ll \R \sim \t^{-1/2}$ appears to be the same as in the Ginsburg-Landau 
theory: $G_{ab}(x-x') = \la \p_{a}(x)\p_{b}(x')\ra \sim |x-x'|^{-(D-2)} \d_{ab}$ 
(beyond $\R$ this correlation function decays exponentially).
Therefore, the typical value of the fluctuations $\la\p^{2}\ra$ can be estimated 
in the usual way:

\be
\lb{g33}
\la\p^{2}\ra \sim \fr{1}{n}\sum_{a=1}^{n} \R^{-D} \int_{|x|<\R} d^{D}x G_{aa}(x) \sim
\t^{(D-2)/2}
\ee
The saddle-point approximation considered in the previous Section is justified
only if the typical value of the fluctuations is small compared to the 
value of the "background" instanton field $\f^{(inst)}(x) \sim \sqrt{\t/\lm(k)}$
(see eq.(\ref{g23})):

\be
\lb{g34}
\t^{(D-2)/2} \ll \fr{\t}{\lm(k)} \; \; \; \Rightarrow \; \; \lm(k) \sim uk \ll \t^{(4-D)/2}
\ee
On the other hand, the contribution (\ref{g29}) appears due to summation in the 
region $k \gg g/u$. Thus, one can get this type of contribution to the free energy 
only in the following interval of summation over $k$:

\be
\lb{g35}
\fr{g}{u} \ll k \ll \fr{1}{u} \t^{(4-D)/2}
\ee
This interval exists provided

\be
\lb{g36}
\t \gg g^{2/(4-D)}
\ee
which is the usual Ginsburg-Landau criteria. 

One can also arrive to the same conclusion deriving 
the fluctuational contribution to the RSB part of the free energy by the direct integration
over fluctuating fields using Gaussian Hamiltonian (\ref{g30}) (this way one can
also check that this contribution contains no imaginary parts which would happen,
if the considered extrema would correspond to physically unstable field configuration).
Assuming the $\theta$-like structure of the instanton solution:
$\psi(|z|\leq 1) \simeq \psi(0) \equiv \psi_{o} \sim 1$ and $\psi(|z| > 1) = 0$,
the fluctuating modes with momenta $p \ll \sqrt{\t}$ and $p \gg \sqrt{\t}$ in the
Hamiltonian (\ref{g30}) can be explicitly decoupled:

\begin{eqnarray}
\lb{g37}
H &=&\2 \sum_{a,b=1}^{n} \int_{|p|\gg\sqrt{\t}} \fr{d^{D}p}{(2\pi)^{D}} 
      \[ p^{2} \d_{ab} + \t T_{ab} \] \p_{a}(p)\p_{b}(-p) +\nonumber
\\ \nonumber
\\
&+&\2 \sum_{a=1}^{n} \int_{|p|\ll\sqrt{\t}} \fr{d^{D}p}{(2\pi)^{D}}  p^{2} |\p_{a}(p)|^{2}
\end{eqnarray}
where $p$-independent matrix $T_{ab}$ is given by eqs.(\ref{g31})-(\ref{g32}),
in which instead of the function $\psi(x\sqrt{\t})$ one has to substitute the constant 
$\psi_{o}$.

The integration over the {\it replica symmetric} modes with momenta $p \ll \sqrt{\t}$
(they corresponds to fluctuations at scales much bigger than the size of the instanton)
described by the second term of the Hamiltonian (\ref{g37}), gives the contribution
of the form $\exp(- n V \tilde f_{RS} )$, and it vanishes in the limit $n\to 0$
(note that in the RSB part of the free energy we have to keep only the terms
which remain {\it finite} in the limit $n\to 0$ and not linear in $n$). This is natural,
because this contribution is already contained in the RS part of the free energy.

The intergation over the modes with momenta $p \gg \sqrt{\t}$ is slightly
cumbersome but straightforward:

\begin{eqnarray}
\lb{g38}
\tilde Z_{RSB} &\equiv& \prod_{p \gg \sqrt{\t}} \[\int {\cal D}\p_{a}(p)\]  
\exp\{-H[\p_{a}(p)] \} =\nonumber
\\ \nonumber
\\
&=&\exp\[ -\2 \t^{-D/2} \int_{p \gg \sqrt{\t}} d^{D}p \Tr \ln \(p^{2}\d_{ab} +\t T_{ab}\) \]
\end{eqnarray}
The matrix under the logarithm in the above equation contains $(k-1)$ eigenvalues:

\be
\lb{g39}
\lm_{1} = p^{2} + \t \(1 - \fr{uk - 3g}{uk - g} \psi_{o}^{2} \)
\ee
one eigenvalue:

\be
\lb{g40}
\lm_{2} = p^{2} + \t \(1 - \fr{uk - 3g}{uk - g} \psi_{o}^{2} \) - \t \fr{2uk}{uk-g} \psi_{o}^{2}
\ee
and $(n-k)$ eigenvalues:

\be
\lb{g41}
\lm_{3} = p^{2} + \t \(1 - \fr{uk}{uk - g} \psi_{o}^{2} \)
\ee
Substituting these eigenvalues into eq.(\ref{g38}), after simple algebra 
in the limit $n\to 0$ one eventually obtains the following result:

\be
\lb{g42}
\tilde Z_{RSB} \sim \exp\(  \fr{3k}{2(uk-g)} g \psi_{o}^{2} \)
\ee
Thus we see that in the region $\t \gg g^{2/(4-D)}$ the factor $kg/(uk-g)$
in the exponential of the above equation is small compared to the 
leading term $k \t^{(4-D)/2}/(uk-g)$ given by the saddle-point solution, eq.(\ref{g25}).

\sectio{Discussion}

It is interesting to note that non-analytic instanton contribution of the form
given by eq.(\ref{g29}) can be easily "derived" basing on qualitative physical
arguments. Let us again consider the random Hamiltonian (\ref{g1}) at temperatures
above $\T$ ($\t > 0$), and let us estimate the contribution to the free energy coming from
rare "ferromagnetic islands" where $\d\t(x) > \t$. In the mean-field regime at finite 
values of $\t$ the typical smallest (most probable) size of such island is
$\R \sim \t^{-1/2}$. Therefore, according to the probability distribution, eq.(\ref{g2}),
in the limit of weak disorder ($u\to 0$) the contribution of the islands to the free energy 
with the exponential accuracy can be estimated by their probability:
\begin{eqnarray}
\lb{g43}
\d F &\sim& \int_{\t}^{\infty} d(\d\t) \exp\( -\fr{(const)}{u} \t^{-D/2} (\d\t)^{2} \) \nonumber
\\ \nonumber
\\
&\sim&\exp\(-\fr{(const)}{u} \t^{(4-D)/2} \)
\end{eqnarray}
which (up to undefined $(const)$ factor) coincides with the result (\ref{g29}).

The above qualitative consideration seems rather valuable because it provides
good physical support for more exact but slightly formal and somewhat mysterious
vector replica symmetry breaking scheme considered in Section 2. 

Of course, exponentially small contributions to the free energy (as well as to others
thermodynamical functions) of the type (\ref{g29}) are not so important for 
thermodynamical properties of the random ferromagnet in the considered
paramagnetic temperature region. Nevertheless, the fact of their existence seems very 
interesting for two reasons. 

First of all, it tells that even in the mean-field regime 
the free energy of the random ferromagnet must be non-analytic function of the 
parameter which describes the strength of disorder $u \to 0$, which is interesting in itself.

Second, it indicates on the importance of non-linear excitations which in terms
of the present replica field theoretical approach are described by the 
localized instanton-like solutions of the stationary equations. In the considered
mean-field region away from $\T$ these excitations provide only exponentially small
corrections. However in the close vicinity of the critical point the presence
of instantons (which is ignored in the standard renormalization-group approach),
and their interactions with the critical fluctuations may produce dramatic effect
on the critical properties of the phase transition. 
It is worthing to note that although in the scaling regime (at $T = \T$) the situation
looks very different from that considered in this paper, the corresponding
stationary equations (\ref{g8}) (with $\t = 0$) also have non-linear instanton-like
solutions with the RSB structure given by eq.(\ref{g14}). One can easily check
that in the dimension $D = 4$  these solutions can be found explicitly \cite{lipatov}:
\be
\lb{g44}
\f(x) = \sqrt{\fr{8}{(uk-g)}} \fr{R}{R^{2} + |x|^{2}}
\ee
where the size of the instanton $R$ appears to be the {\it zero mode}
(the energy of the instanton does not depend on $R$). In dimensions
below but close to four (at $\e = (4-D) \ll 1$) the field configuration given by
eq.(\ref{g44}) can be considered as the approximate solution  which contains the 
parameter $R$ as the {\it soft mode}, since the energy of the instanton, eq.(\ref{g16}), 
depends on $R$ very weakly:
\be
\lb{g45}
E(k) = \fr{4}{3} S_{D} R^{-\e} \fr{k}{uk-g}
\ee
(here $S_{D}$ is the square of the unite $D$-dimensional sphere).

At present it is not qute clear how all these non-linear instanton excitations
could be incorporated into the self-consistent theory of the critical fluctuations.
Keeping in mind that the degrees of freedom of this type explicitly break the replica 
symmetry, a kind of "heuristic" renormalization group approach has been 
proposed \cite{d-rsb}, in which it was assumed that due to interactions of the
fluctuations with this type of non-perturbative excitations the replica symmetry in the 
effective matrix, describing non-linear interactions of the fluctuating fields, is 
spontaneously broken. This resulted in the the instability of previously known
fixed points and remarkable  "runaway" behaviour of the renormalization group
flows (which e.g. may indicate on  the onset of a kind of the glass-like phase
in a narrow temperature interval around $\T$). 
I hope the study described in the present paper would stimulate further much deeper
investigation of the physics of the phase transition in random ferromagnets.

\vspace{5mm}

{\bf Acknowledgements}

\vspace{2mm}

The author is greatful to M.M\'ezard, Vl.Dotsenko, G.Parisi and S.Franz for useful 
discussions.

\end{document}